\documentclass[12pt]{article}

\newtheorem{Theorem}{Theorem}[section]
\newtheorem{Definition}[Theorem]{Definition}
\newtheorem{Proposition}[Theorem]{Proposition}
\newtheorem{Lemma}[Theorem]{Lemma}

\makeatletter
\usepackage{latexsym}
\usepackage[T1]{fontenc}
\usepackage{amsmath}
\usepackage{amssymb}

\newtheorem{Corollary}[Theorem]{Corollary}

\def \AO {{\cal A}({\cal O})}
\def \AO' {{\cal A}({\cal O}')}

\def \] {\supseteq}

\def \Pf {{\bf Proof.\,\,}}

\def \be {\begin{equation}}
\def \ee {\end{equation}}

\def \ume {{\scriptstyle{\frac{1}{2}}}}

\def \eqq {\equiv}

\def \a {{\alpha}}

\def \A {{\cal A}}
\def \B {{\cal B}}
\def \C {{\cal C}}

\def \G {{\cal G}}
\def \H {\mbox{${\cal H}$}}

\def \L {{\cal L}}

\def \O {{\cal O}}

\def \S {{\cal S}}
\def \T {{\cal T}}
\def \U {{\cal U}}

\def \Z {{\cal Z}}

\def \id {{\bf 1 }}

\def \Rbf {{\bf R}}

\def \AO {{\cal A}({\cal O})}
\def \AO' {{\cal A}({\cal O}')}

\def \] {\supseteq}

\def \Pf {{\bf Proof.\,\,}}

\newfam\Ssfam
\font\eleSs=cmss10 at12pt \font\sevenSs= cmss10 at 8pt \font\sixSs= cmss10 at 6pt

\textfont\Ssfam=\eleSs\scriptfont\Ssfam=\sevenSs%
\scriptscriptfont\Ssfam=\sixSs \def\Ss{\fam\Ssfam\eleSs}

\def\doppio#1{{\rm I}\kern-.1667em{\rm #1}}

\def\R{{\cal R}}
\def\Q{\text{Q}\kern-.52em
    \text{\vrule height1.5ex width.5pt depth0pt}\kern.45em}

\def\Z{{\mathchoice {\hbox{$\Ss\textstyle Z\kern-0.4em Z$}}
{\hbox{$\Ss\textstyle Z\kern-0.4em Z$}} {\hbox{$\Ss\scriptstyle Z\kern-0.25em
Z$}} {\hbox{$\Ss\scriptscriptstyle Z\kern-0.2em Z$}}}}

\def\C{{\mathchoice{\hbox{$\rm\textstyle\text{\kern.35em\vrule
   height1.5ex width.5pt depth0pt\kern-.35em C}$}}
{\hbox{$\rm\textstyle\text{\kern.35em\vrule
   height1.5ex width.5pt depth0pt\kern-.35em C}$}}
{\hbox{$\rm\scriptstyle\text{\kern.35em\vrule
   height1.5ex width.3pt depth0pt\kern-.35em C}$}}
{\hbox{$\rm\scriptscriptstyle\text{\kern.35em\vrule
   height1.5ex width.2pt depth0pt\kern-.35em C}$}}}}

\def \be{\begin{equation} \displaystyle}
\def \ee{\end{equation}}

\def \A*{\mbox{$A^{*} $}}
\def \B*{\mbox{$B^{*} $}}
\def \C*{\mbox{$C^{*} $}}

\def \id{\mbox{${\bf 1}\,$}}

\def \bea{\begin{eqnarray}}
\def \eea{\end{eqnarray}}

\def \Pf{{\em Proof.\,\,}}

\def \a{\alpha}

\@addtoreset{equation}{section}

\def \be {\begin{equation} \displaystyle}

\def \ee {\end{equation}}

\def\AO {\mbox{${\cal A}({\cal O})$}}
\def\AO'{\mbox{${\cal A}({\cal O}')$}}
\def\O {\mbox{${\cal O}$}}

\def\A{\mbox{${\cal A}$}}

\def \O {{\cal O}}

\def \A {{\cal A}}
\def \AO {\A(\O)}
\def \AOl'{\A(\O_{loc}')}

\def \B {{\cal B}}

\def \H {{\cal H}}

\def \S {{\cal S}}

\def \M {{\cal M}}

\def \difm  { \mbox{Diff}(\M)}

\def \cmo {{C_0^\infty(\M)}}
\def \cif {{C^\infty}}

\def \cg {{C_g}}
\def \vg {{V_g}}
\def \vh {{V_h}}

\begin{document}
\begin{titlepage}
  \title{Manifold Topology, Observables and Gauge Group}

\sloppy

\author{G. Morchio and F. Strocchi \\Dipartimento di Fisica, Universit\`a di Pisa,
%\\and  INFN, Sezione di Pisa,
 Pisa, Italy  }

\fussy

\date{}

\maketitle

\begin{abstract}

The relation between manifold topology, observables and gauge group is clarified
on the basis of the classification of the representations of the algebra of
observables associated to positions and 
displacements on the manifold.
The guiding, physically motivated, principles are
i) locality, i.e. the generating role of the algebras localized in small, topological trivial, regions,
ii) diffeomorphism covariance, which guarantees the intrinsic character of the analysis,
iii) the exclusion of additional local degrees of freedom with respect to the Schroedinger representation.
The locally normal representations of the resulting observable algebra are
classified by 
unitary representations of the fundamental group of the manifold,
which actually
generate an observable, ``topological'', subalgebra.
The result is confronted with the standard approach based on the introduction
of the universal covering ${\tilde{\cal M}}$ of $\cal{M}$ and on the
decomposition of $L^2({\tilde{\cal M}})$ according to the spectrum of the fundamental group, 
which plays the role of  a gauge group.
It is shown that in this way one obtains all the representations of the observables iff the
fundamental group is amenable. The implications on the 
observability of the Permutation Group in Particle Statistics are discussed.

\end{abstract}

\vspace{15mm} \noindent
Math. Sub. Class.: 81Q70, 81R15, 81R10

\vspace{3mm}  \noindent

\end{titlepage}

%%%%%%%%%%%%%%%%%%%%%%%%%%%%%%%%%%%%%%%%%%%%%%%%%%%%%%%

\newpage
\section{Introduction}

For the analysis of quantum mechanical systems with
a non-trivial manifold $\M$ as configuration space, the
r\^ole of the topology of $\M$ has been investigated by many authors
\cite{Gol}, \cite{Do}, \cite{Du} \cite{Sch}, \cite{La}, \cite{La1},
\cite{noi}.
The involved fundamental issues include the control of the representations of the
Diffeomorphisms Group of the manifold \cite{Mil}, the identification
of the observable algebras and the classification of their representations.
Similar problems arise for the formulation of
Local Quantum (Field) Theory on space-time manifolds \cite{Bru}.

Historically, the main approaches to Quantum Mechanics (QM) on manifolds,
the Functional Integral formulation \cite{DeW, Sch}
and Geometric Quantization (GQ) of symplectic manifolds \cite{Sor, Sor1, Mor, Rief},
produced substantially equivalent results.
Apart from the possible presence  of an ``internal'' space, such methods
lead in fact to Schroedinger QM in $L^2(\tilde\M)$, $\tilde\M$ the universal covering space of $\M$,
with $\pi_1(\M)$ playing the r\^ole of a gauge group.
The centre of the representation of the gauge group in $L^2(\tilde\M)$
is observable and reduces $L^2(\tilde\M)$ to inequivalent Quantum Mechanical ``sectors'', 
with a multiplicity given by the action of the gauge group.

The result can be seen as a generalization of Dirac treatment of
Identical Particles \cite{Dirac}, which identifies the observable algebra
through the invariance under the  Permutation Group, which therefore plays the role of a gauge group, 
and classifies its representations by those of the gauge group.
In three space dimensions, the relation with Dirac treatment of identical particles
is indeed very close, since the Permutation Group
coincides with the fundamental group of the configuration manifold of identical
particles \cite{LaStat}.

In view of their implications on basic issues, however,
the above treatments leave open substantial questions.
In fact,  they rely  on
``quantization'' prescriptions, associating a quantum system to a ``corresponding''
classical one, without an independent examination the physical basis and interpretation of
the adopted approach.
The situation is therefore quite different with respect to the case of QM
in $\Rbf^d$, where the approach based on
 Weyl algebras provides a clear and unique result, through the Von Neumann
uniqueness theorem.

In particular, the question arises whether an observable role is associated to the
entire group $\pi_1(\M)$, or 
only to its center, as indicated by Dirac analysis. 
To this purpose, clearly, the identification of  the
QM observables on the basis of physical principles becomes the decisive issue, also 
because no comparable phenomena appear for classical particles.

The need of a description in terms of observables and states has been
emphasized by Landsman \cite{La1, LaStat}.
His approach is in fact based on the construction of an observable algebra,
which is obtained by ``Deformation Quantization'' on
$\tilde\M$, taking a quotient with respect to the action (on $\tilde \M$) of $\pi_1(\M)$, which
 therefore plays the role of a gauge group.
In Landsman's analysis, the same group 
also appears as a unitary group in the observable algebra,
and \emph {in such a role} it provides the classification of its representations.
Thus, on one side Landsman's treatment follows the approach and confirms the conclusions
of the previous literature:
\emph {the fundamental group of the manifold plays the role of a gauge group}, 
which identifies   the observable algebra starting from $\tilde \M $; on the other, Landsman's
classification of the irreducible representations of the observable algebra \emph {does not}
involve the gauge group, but rather an isomorphic \emph {observable} group.
Clearly, two questions arise:

First, since $\tilde\M$ is taken as a starting point in \emph {all} the above approches,
one may ask whether the above results depends on such a choice, which does
not seem to have a direct physical interpretation. In any case, as argued above, a
direct foundation on physical principles is lacking.

Second, if $\tilde \M$ needs not to appear in the formulation,
the classification of the representations in terms of 
$\pi_1(\M)$ as a gauge group is in question.
Moreover, even if the different formulations can be compared,
the classification in terms of $\pi_1(\M)$ as 
a gauge group and as an observable group are not a priori related and
to which extent they may coincide is an open problem.

An answer to the first question has been given in ref. \cite{noi}, which
avoids the introduction of $\tilde\M$ and directly identifies the observable algebra
as generated by coordinates and momentum variables associated
to vector fields on $\M$.
The construction only employs the manifold $\M$, the identification of
the observables does not make any reference to $\tilde \M$ and no action of $\pi_1(\M)$
as a gauge group appears.
The irreducible representations of the observable algebra 
are then classified by a representations of $\pi_1(\M)$, which appears
\emph {within each representation space}.
By irreducibility of the observables in each sector, such a representation of
$\pi_1(\M)$ is automatically given by strong limits of observables.

The purpose of this note is twofold: first, to construct the observable algebra on the basis 
of clear  physical principles, holding independently of the topology of the manifold.
The identification of the observable algebra will be 
performed purely in terms of a collection of \emph {local} algebras,
localized in ``small'', topologically trivial, regions.
This also allows for a purely local analysis
of the degrees of freedom which may appear, as a consequence of
diffeomorphism invariance, in addition to Schroedinger QM.
The local and global effects are therefore separated completely,
and the role of the topology precisely appears when the local
(essentially Schroedinger) descriptions are glued together. 

The second aim is to confront the resulting representations,
classified by an \emph{observable} $\pi_1(\M)$, with those obtained by
the Dirac gauge group method.

\vspace{2mm}

\noindent
{\bf Outline of the strategy and  main results}

\smallskip
\noindent
The Observable Algebra for the description of a
quantum particle on a $d$-dimensional manifold $\M$ is constructed
according to the following physically motivated principles
and with the following results:

\vspace{2mm}
\noindent
1. {\bf \em The local structure}

\smallskip
\noindent
a) \emph{Locality}: The observable algebra is generated by a collection of local algebras associated to
(topologically trivial) ``small'' regions $\O$, technically defined as regions
homeomorphic to a subdisk of an open disk (homeomorphism to an open disk would
not be enough to exclude, e.g., the entire space $\Rbf^d$). 

\smallskip
\noindent
b) \emph{Diffeomorphism Covariance}: The identification of the observables is exclusively
linked to the manifold structure.
Locally, it amounts to  independence of coordinates, as well as of a metric,
special transformation groups etc.; globally,
it provides the intrinsic geometric links between the
observable algebras localized in different regions, just as Poincar\'e covariance
does in relativistic quantum field theory.  

\smallskip
\noindent
c) \emph{Positions and "local Movements" on the manifold}: The local algebras are generated by position
and ``local trasportation'' variables, the  latter identified with the unitary groups $\G(\O)$
describing displacements along (all, localized) vector fields with support in $\O$.

\smallskip

The local algebras are therefore assumed to be generated
by $C^0$ functions $\alpha(x)$ with support in $\O$,
as position variables, and by unitaries $U(g)$ representing the diffeomorphisms groups
$\G(\O)$. They are identified with the Crossed Products $ \Pi(\O) \equiv C^0(\O) \times \G(\O)$,
defined by the relation
$ U(g) \, \a(x) \, U(g)^* =\a(g^{-1} x) \, $,
representing the action of diffeomorphisms $g \in \G(\O)$ on functions on $\O$.

The algebras $\Pi(\O)$ generate
$\Pi(\M) \equiv C^0(\M) \times \G_L(\M)$, as global observable algebra,
with $\G_L(\M)$ the group generated by all the $\G(\O)$
(see Section 2.3).
No independent global variables are therefore introduced and the topology of $\M$ only
affects the result of algebraic operations on local observables.

\vspace{3mm} 

\noindent
2. {\bf \em The Local One (Spinless) Particle Interpretation}

\smallskip

\noindent
As a consequence of diffeomorphism invariance,  the local algebras 
 admit many inequivalent representations (also describing many particle systems);
hence, for a one-particle description, a selection is needed to avoid, at the local level, the presence
of additional degrees of freedom, with respect
to those appearing in Schroedinger QM.
Spin and internal degrees of freedom, excluded by the above requirement, may be added by an explicit
extension of the observable algebra.

The representation of the local algebras can be taken in spaces of the form
$$ \H(\O) = L^2(\O, d\mu) \times K(\O)\, ,$$
(see Sect.2 for the exact characterization),
with the association of unitary ``cocycles'' 
$V_g(x)$
in $K(\O)$, $g \in \G(\O)$, $x \in \O$, to local displacements.

Up to unitary equivalence, such cocycles are characterized by 
those representing the subgroup  $\G(\O, x)  \subset \G(\O)$ which leaves a point $x \in \O$ stable;
the unitary class of their representations is independent from the point and the region.

Such cocycles, describing modifications of local displacements by operations which leave the
point invariant, have the physical interpretation of describing ``internal degrees of freedom''.
\emph {Their triviality}, which excludes such degrees
of freedom, is required by a \emph {Local One Particle Interpretation}
and turns out to be equivalent to the Schroedinger representation of the local
algebras $\Pi(\O)$, up to a multiplicity ({\em Locally Schroedinger  condition}). 

The exclusion of internal degrees of freedom is actually a general question
for representations of observable algebra which include momenta 
 describing  ''movements'' besides
those associated to translations. In fact, it appears  already in the case
of one spinless quantum particle in $\Rbf^d$ with momenta indexed by the generators of
the euclidean group, where non-trivial cocycles are associated to rotations and describe spin;
more generally, the same problem appears for the representation of the stability group
of a point in the case of the algebra constructed from a group $G$ and associated to the manifold
$G/H$, $H$ a subgroup of $G$ \cite{LaGrup}.

As a result, on the basis of the above principles, we 
obtain a complete identification
of the local observable algebras and of their representations;
\emph {they simply describe Schroedinger QM, up to a multiplicity,
  in diffeomorphism covariant variables}.

The control of QM of a particle on a manifold is then reduced to
the classification of the representations of the \emph {global} algebra
$\Pi(\M) \equiv C^0(\M) \times \G_L(\M)$
which reduce to a multiple of the Schroedinger representation when restricted
to $\Pi(\O)$, for all $\O$.

Equivalently, in the spirit of Local Quantum Theory \cite{Haag}, the local observable algebras
can be identified with the weak closures $\A(\O)$ of $\Pi(\O)$
in (any) LS representations of $\Pi(\M)$. The global
observable algebra for one particle on $\M$ is then identified
with the algebra $\A(\M)$ generated by them in the sum of such representations.
LS representations of $\Pi(\M)$ coincide with
locally normal representations of $\A(\M)$. 

\vspace{2mm}

\newpage
\noindent
3. { \bf \em Classification of Locally Schroedinger (LS) representations}

\smallskip

\noindent
 By the above results, the classification of the LS representations of $\Pi(\M)$  
 is reduced to the analysis of the cocycles $V_{g_n \ldots g_1} (x)$ associated to products of
 localized diffeomorphisms $g_i \in \G(\O_i)$.
 They are shown to depend only on the topological equivalence class of the path
from $x$ to $g_n \ldots g_1 x$ 
given by $g_n \ldots g_1$, and 
to define a unitary representation of the path grupoid, which is classified,
up to unitary equivalence, by its restriction to closed paths
with (any) fixed base point, i.e., by a representation of $\pi_1(\M)$.

 All unitary representations of $\pi_1(\M)$ are shown to define admissible cocycles, so that
the correspondence between LS representations of $\Pi(\M)$
and unitary representations of $\pi_1(\M)$ is one to one.
If $M$ is simply connected, 
the \emph { analogue of  Von Neumann uniqueness theorem} holds:
the (locally normal) representation of 
$\A(\M)$ is \emph {unique} (and Schroedinger), up to a multiplicity. 

The above representations of $\pi_1(\M)$, multiplied by projections over small regions $\O$,
are shown to describe observables corresponding to the transport of the particle,
starting in the region $\O$, along the corresponding loops.
Moreover, for compact $\M$, $\A(\M)$
contains a subalgebra isomorphic to the group algebra of $\pi_1(\M)$.
The classification of the representations is therefore always given by
an \emph {observable} representation of $\pi_1(\M)$.

\vspace{2mm}

\noindent
4. {\bf \em The gauge and observable realizations of $\pi_1(\M)$}

\smallskip

\noindent
 A comparison with the Dirac gauge group approach \cite{Dirac} is provided by the analysis of the
Schroedinger representation of $\A(\M)$ in $L^2(\tilde \M)$.
Identifying $\tilde \M$ with a space of pairs
$x \in \M , \gamma \in \pi_1(\M) $, one has
$$ L^2(\tilde \M) \sim L^2(\M) \times l^2(\pi_1(\M)) \, .$$
The usual ``gauge'' action of $\pi_1(\M) $ in $\tilde\M$
is given by
$$ \psi (x,\gamma) \to \psi (x,\gamma \circ \delta) \, ,$$
i.e., by its \emph {right} regular representation in $l^2(\pi_1(\M))$.
It clearly commutes with the position observables and also
%19 ho tolto ``commutes'' e ho cambiato sotto ``whose'' in ``which...''
with the unitaries $U(g)$, 
which act as in the Schroedinger representation in 
$L^2(\M)$, combined with the \emph {left} regular representation of $\pi_1(\M))$ in
$ l^2(\pi_1(\M))$ (see eq.\,3.7).

 It follows that, in the representation of $\A(\M)$ in $L^2(\tilde \M)$,
\emph {two} commuting unitary representations of $\pi_1(\M))$ appear,
respectively as a gauge group \emph {and} as an observable group.
Being given by (a multiple of) the left and right regular representations,
 the two representations
 are unitarily equivalent; they generate Von Neumann algebras which are the
commutant one of the other in $ l^2(\pi_1(\M))$, so that
their centres coincide and give rise to \emph {the same reduction of} 
$L^2(\tilde \M)$. 
Therefore, the classifications of the representations of $\A(\M)$ in $L^2(\tilde \M)$
given by the observable and by the gauge group 
 \emph {coincide}.

 As a result, only the completeness of the Dirac approach is in question, i.e. whether
\emph{all} the (locally normal) irreducible representations of $\A(\M)$
appear in the reduction of $L^2(\tilde \M)$.
Equivalently, whether 
\emph {all} the irreducible representations of $\pi_1(\M)$
appear in the (possibly integral) reduction of its regular representation.
The answer to this questios is  known and rather simple: this happens if and only if
the group is \emph {amenable}. 

 If $\pi_1(\M)$ is not amenable, the role of its regular representation and of $L^2(\tilde \M)$
is completely different, since even the identity representation of $\pi_1(\M)$, corresponding
to the ordinary Schroedinger representation of the observables, is \emph {not} present
in the reduction.
Proposition 3.3 below shows that in this case all the representations of $\A(\M)$
can still by obtained in suitable Hilbert spaces of functions on $\tilde \M$ (with non $L^2$
scalar products), but the role of the gauge group is in general lost.

\vspace{2mm}

\noindent
5. {\bf \em Implications for Identical Particles}

\smallskip
\noindent
Particularly significant is the case of $N$ identical particles,
which can be described as the quantum system associated to the $N$-particle
manifold $\M_S$, defined by identifying configurations obtained
by permuting the particles and excluding the set $\Delta$ of coincident points
\cite{Leinaas}.

In the case of $N$
particles in Euclidean space of dimension 
greater than two, the fundamental group of $\M_S$
is the permutation group $S_N$ 
and $\tilde\M_S$ may be identified with $ \Rbf^{Nd} \setminus \Delta $.
From the above results it follows that

\noindent
i)  since $S_N$ is finite (and therefore amenable), the Dirac representation of
the observable algebra in $L^2(\Rbf^{Nd}) $ ($\Delta$ being here irrelevant)
contains all its irreducible representations, classified by the
representations of $S_N$ in $L^2(\Rbf^{Nd}) $ as a gauge group.

\noindent
ii) an observable unitary representations of the Permutation Group
is present in $L^2(\Rbf^{Nd})$, unitarily equivalent to the gauge representation.
In each reduction space, the gauge and observable representations 
are complex conjugates.

The observable representation acts (see Sect.\,3) by \emph {physical operations
which shift the position of each particle} (as a gauge invariant variable,
independent of particle labels), along paths which interchange the position of the particles.
Apart from the abelian case, $N = 2$, the (gauge invariant) operators
implementing such actions have nothing to do with gauge trasformations,
which in fact act \emph {on the particle labels} and are not observable.

Clearly, the presence of the observable representation 
of $S_N$ allows for a direct
physical (and topological) reinterpretation of the Dirac classification,
in terms of unitaries describing permutations as physical operations.
\goodbreak

\section{The observable algebra and its Locally Schroe\-dinger representations}

\subsection{The algebra generated by local coordinates and local displacements}

In the standard case of a quantum mechanical
system with $\Rbf^d$ as configuration space, the 
observables are usually generated by the Cartesian coordinates in $\Rbf^d$
and the associated  momenta;
Schroedinger Quantum Mechanics then arises as the unique (regular) representation 
of the Weyl algebra, generated by the exponentials of such variables.
On the other hand, when the configurations are described by a
$\cif$ connected ($d$-dimensional) manifold $\M$
no global coordinate system is available
and, even locally, there
are no distinguished coordinate systems, nor 
intrinsic finite dimensional transformation groups.

Following the general philosophy of Local Quantum Theory \cite{Haag},
we start by considering ``local'' open regions $\O $, 
{\em topologically trivial and
topologically localizable}, in the sense that they are proper subsets of larger similar
regions. In the following, $\M$ will indicate any 
$\cif$ connected manifold.

\begin{Definition}
An open region $\O \subset \M$ is called {\bf \em small} if there is 
a diffeomorphism of a neighbourhood $\O'$ of $\O$
taking $\O'$ to an open ball, $|x| < 1$, $x \in \Rbf^d$,
and $\O$ to the open ball $|x| < 1/2$. 
\end{Definition}

Hereafter, $\O $ will always denote a small region.
Clearly, each point of $\M$ has a small open neighbourhood $\O$
and a denumerable set of such regions covers $\M$.

Our guiding principle for the construction of the local observable algebras
is to recognize as fundamental the
covariance under diffeomorphisms, 
linking the observables to the intrinsic geometry of $\M$.
For each $\O $ we introduce: 

\noindent
i) as ``{\em position observables}'' the $\cif $ (complex valued) functions
$\a(x)$, $x \in \M$, with compact support in $\O$, 
generating, with the Sup norm and together with a common identity $\id$,
a diffeomorphism invariant $C^*$ algebra $C(\O)$;
then, $C(\O)$ is isomorphic to the algebra generated by
the continuous functions on $\O$, vanishing at its boundary, and by
the  constant functions. 

\noindent
ii) as ``{\em local generalized momenta}'', the vector fields  $ v \in \L(\O)$,
$\L(\O)$ the Lie algebra of $\cif$ vector fields with compact support in $\O$;
by compactness of their support, they integrate to one parameter groups $g_{\lambda v}$,
 $\lambda \in \Rbf$, generating diffeomorphism groups $\G(\O)$,
with a common identity $ \mathbf e$ and
$\G(\O_1) \subset \G(\O_2)$ for $\O_1 \subset \O_2$. 
In physical terms, $\G(\O)$ describes displacements \emph {localized in} $\O$,
acting on (any) configuration of the system.

\noindent
The vector space of the pairs $\{\alpha, \,g\,\}, \alpha \in C(\O), \,g \in \G(\O)$,
with the operations 
$$ \{\alpha_1, \,g_1\,\}\,\{\alpha_2, \,g_2\,\} = \{\alpha_1\,
\alpha_2^{g_1} , \,g_1\,g_2\,\}, \,\,\,\, \alpha_2^{g}  (x) \eqq \alpha_2 (g^{-1} x)
\,  \ \ \{\alpha, \,g \,\}^{*} = \{\bar \alpha, \,g^{-1} \,\} \, ,$$
defines the 
{\bf \em  crossed product $C^*$ algebra}
$ \Pi(\O) \equiv C(\O) \times \G(\O)$. 

In the construction of the crossed product, $\G(\O)$ is
taken as a topological group with the discrete topology and
$\Pi(\O)$ is generated by the finite sums
$ \sum_i \alpha_i \, g_i $, $\alpha_i \in C(\O)$, $g_i \in \G(\O)$,
with norm $ \sum_i {\rm Sup}_x |\alpha_i(x)|$, see, e.g.,
\cite{CrossPr}.
In the following, for simplicity, we adopt the 
% 21' ho tolto "following", che non è indispensabile, riottentedo così la precedente impaginazione 
notations: 
$U(g) \eqq \{ \id ,\, g\,\}, \, \a(x) \eqq \{\alpha, \,\mathbf e \}$;
then the basic crossed product algebraic relations read  
\be 
\label{crossed}
U(g)^{-1} =  U(g)^* \,  , \   
U(g) \, \a(x) \, U(g)^* =\a(g^{-1} x) \,  ,  \
{\rm {Supp}} \, \alpha \subset \O \, , \,\,g  \in \G(\O) \, .
 \ee
The {\em  local ``position'' algebras} $C(\O)$ satisfy isotony,
$C(\O_1) \subset C(\O_2)$ for $\O_1 \subset \O_2$
and generate, in the Sup norm on $\M$
($\id$ being identified with the function $1$ on $\M$),
the $C^*$ algebra $ C(\M) $ of continuous functions on $\M$,   
if $\M$ is compact, and on its one-point compactification
$\dot{\M} = \M \cup \{x_\infty\}$ 
(the Gelfand spectrum of $C(\M)$) otherwise.
In the latter case, the diffeomorphisms of $\G(\O)$
extend to diffeomorphisms of $\dot{\M}$, with
$g x_\infty = x_\infty$; $\alpha(x_\infty) \equiv 0$
for $\alpha(x)$ in $\cif(\O)$, for all $\O$.

The {\em local groups} $\G(\O)$ coincide with the connected component of the identity
of the diffeomorphism group of $\O$ (see \cite{Mil}).
They obviously satisfy isotony and generate a group $\G_{L}(\M)$, uniquely associated to $\M$,
defined by the formal products of their elements modulo the group relations holding in 
each region $\O$.
Its elements are therefore strings $g_1 \ldots g_n$, modulo the equivalence relation defined
by any sequence of replacements of a substring of elements localized in some $\O$ by another
string with the same product in $\G(\O)$.
$\G_{L}(\M)$ 
acts on $\M$ by diffeomorphisms (depending in fact only on the equivalence class
of $g_1 \ldots g_n$),  
which will be denoted by the same symbol. 
$\G_{L}(\M)$ formalizes, in the construction of the observable algebra,
the principle that \emph {finite sequences of local operations still define physical operations
  and are only constrained by the validity of all local relations}. 

We therefore associate to $\M$ the {\bf \em  crossed product $C^*$ algebra}
$\Pi(\M) \equiv C(\M) \times \G_L(\M)$, defined as above, with
eqs.(\ref{crossed}) satisfied for
all $\alpha \in C(\M)$ and $ g \in \G_L(\M)$.
$\Pi(\M)$ is generated by the local algebras $\Pi(\O)$ and 
is invariant under $\difm$, the entire diffeomorphism group
of $\M$, since both $C(\M)$ and $\G_L(\M) $ are invariant.

As we shall see, diffeomorphism invariance makes
such algebras, already for regions $\O$, much richer than, e.g.,
Weyl algebras in $\Rbf^d$; in fact, they admit many inequivalent
representations, with different physical interpretation.
This happens because 
additional independent ``momentum'' variables appear, as a consequence of
diffeomorphism invariance. 

\subsection{Regular representations and their cocycles}
In order to classify the physically relevant representations of $\Pi(\M)$,
we start by characterizing, under physically motivated  conditions,
the representations of its local subalgebras $\Pi(\O)$, which generate it.

First, in order to ensure the existence of local momenta,
we consider representations of $\Pi(\M)$ in separable Hilbert spaces,
satisfying, for each $\O$,  \emph {strong continuity} in $\lambda$ of the local
groups $U(g_{\lambda v})$, $v \in \L(\O)$.

Then, the  representations of $\Pi(\M)$, as well as of $\Pi(\O)$,
are characterized by the following Proposition, which applies to a generic manifold;
for economy of notation, it is stated for $\Pi(\M)$.

\begin{Proposition}
  A representation  $\pi$ of $\Pi(\M)$ in a separable Hilbert space,
  with $U(g_{\lambda v})$ strongly continuous in $\lambda$
  ({\bf \emph {regular representation}})
  is unitarily equivalent to one in 
 \be { \H_\pi = L^2(\M, d \mu)\times K \oplus K_\infty \, ,}\ee
    $K$ and $K_\infty$ separable Hilbert spaces,
  $d \mu$ equivalent to the Lebesgue measure on $\M$
  (in any system of local coordinates).

\noindent 
Identifying the elements of $\H_\pi$ with $L^2$ functions $\psi(x)$,
  $x \in \dot{\M}$, 
  taking values in $K$ for $x\in \M$ and in $K_\infty$ for $x = x_\infty$, the 
action of the representatives of the elements of $\Pi(\M)$ is given by: 

\be{\pi(\a)
  \psi(x) = \a(x) \psi(x), \,\,\,\,\,\,\,\pi(U(g)) = C_g V_g, } \label{Ug}
\ee
\be{ \cg \psi(x) \eqq \psi(g^{-1} x) \, J_g(x)^{1/2}, \,\,\, 
  \ \ \ \
  \vg \psi(x) = \vg(x) \psi(x) \, ,}  \label {Cg}
\ee
$J_g(x) \equiv [d \mu(g^{-1} x)/ d \mu(x)]$,
$\vg(x)$ a family of unitary operators, in $K$ for $x \in \M$ and in
$K_\infty$ for $x = x_\infty$,
weakly measurable in $x$, satisfying
\be{ C^{-1}_h\,V_g(x)\,C_h = V_g(hx),}\ee
\be{\vg(h x) \vh(x) = V_{gh}(x) \, ,} \label{cocycle} \ee
a.e. in $x \in \M$. 

\noindent
Two (regular) representations
$\pi_1$, $\pi_2$ of $\Pi(\M)$ in $L^2(\M, d \mu)\times K_i \oplus K_{\infty, i}, \, i = 1, 2,$
are unitarily equivalent iff there exists a weakly
measurable family of isometric operators $S(x)$ from $K_1$ to $ K_2$ for $x \in \M$, and
from $K_{\infty, 1}$ to $ K_{\infty, 2}$ for $x = x_\infty$,
such that
\be \label{UE} 
     S(gx)\,V_g^{(1)}(x)\,S(x)^{-1} = V_g^{(2)}(x). 
      \ee

\end{Proposition}

The proof is essentially the same as in \cite{noi}, Lemma 3,
only strong continuity of $U(g_{\lambda v})$ for $v \in \L(\O)$
being required; the irreducibility condition is replaced by separability of $\H_\pi$.
The compactification of $\M$ arises as the Gelfand spectrum
of $C(\M)$ and gives rise in general to a representation of
$\Pi(\M)$, in $K_\infty$, which assignes the null value to all
functions with compact support and reduces to a unitary
representation of $\G_L(\M)$ by $\vg(x_\infty)$.

Thus, up to unitary equivalence, the regular representations of the local algebras $\Pi(\O)$
are given  in $ L^2(\O, d \mu)\times K(\O) \oplus K_\infty(\O) $
by eqs.(\ref{Ug}),(\ref{Cg}), with  $\alpha \in C(\O)$,  $g \in \G(\O)$.
Given a regular representation of $\Pi(\M)$, by Proposition 2.2,
the corresponding representation spaces are $K(\O)= K$ for all $\O$ and
$$ K_\infty(\O) = L^2(\M \setminus \O, d \mu)\times K \oplus K_\infty \, .$$

\vspace{2mm}

By Proposition 2.2, the classification of the representations of $\Pi(\M)$
reduces to that of the unitary operators $V_g(x)$.
For $V_g(x) \eqq \id$ and $K$ one dimensional, the representation 
in $L^2(\M, d \mu)$ defines the
{\bf \em Schroedinger representation} $\pi_S(\Pi(\M))$.

Eq.(\ref{cocycle}) becomes
a \emph {group cocycle relation} if the excluded zero measure subset
can be taken independent of $g$ and  $h$.
In fact, in this case, the operators $V_g(x)$ are well defined
as maps from $g \in \G_L(\M)$ to the group $\U$
of unitaries in $K$ depending on $x \in \M \setminus A$, $A$ of zero measure; 
eq.(\ref{cocycle}) is then the cocycle relation associated to the map
$$ g \to \varphi(g) \, , \ \ \ \ \mbox {Aut} (\U) \ni \varphi(g):  V(x) \to V(gx) \, . $$ 

\begin{Definition}
  A representation $\pi$ of $\Pi(\M)$ will be called {\bf \emph {cocycle-regular}} if
  it is unitarily equivalent to a regular representation where
  eqs.(\ref{cocycle}) define a group cocycle relation,
  for almost all $x$ in $\M$.
\end{Definition}  

In a cocycle-regular representation of $\Pi(\M)$
the operators $V_g(x)$ provide a representation of
the \emph{stability groups} 
$$ \G(\O ,x) \eqq \{ g \in \G (\O) : g x = x \} \, , \ \ \ x \in \O \,  $$
in $K$, for almost all $x \in \M$.

For different $\O \subset \M$, $x \in \O$, the groups $\G(\O)$ are isomorphic,
and the same applies to $\G(\O,x)$.
In fact, it is enough to consider  $\O_1$, $\O_2$ disjoint, $x_i \in \O_i$; then,
there exists a region $\O \supset (\O_1 \cup \O_2)$  and a diffeomorphism  $g \in \G(\O)$
transforming  $\O_1$ in $\O_2$ and $x_1$ in $ x_2$, constructed, e.g., by
contracting $\O_i$ to sufficiently small balls around $x_i$, interpolated by local
translations in a cylinder around a path from $x_1$ to $x_2$.

The analysis of local cocycles allows for a characterization of the
representations of the local algebras.

\begin{Proposition} 
A cocycle-regular representation $\pi$ of $\Pi(\O)$,
by Proposition 2.2 in $ L^2(\O, d \mu)\times K(\O) \oplus K_\infty(\O) \,$,
defines unitary representations $R(\O,x)$ of $ \G(\O,x)$ in $K(\O)$,
for almost all $x \in \O$, all unitarily equivalent
and a unitary representation $R_\infty$ of $\G(\O)$ in $K_\infty(\O)$.
The corresponding unitary equivalence classes 
determine the cocycles
$V_g(x)$, and therefore $\pi(\Pi(\O))$, up to unitary equivalence.

\noindent
Conversely, any pair of unitary representations of $ \G(\O, x)$ and $ \G(\O)$, respectively
$R, R_\infty$, in spaces $K, K_\infty$
(strongly continous in the parameters of one-dimensional subgroups)
determines a cocycle-regular representation $\pi_{R, R_\infty}$
of $\Pi(\O)$, in $L^2(\O, d \mu)\times K \oplus K_\infty $.
\end{Proposition}

\noindent
\Pf 
By definition of small regions, there exist
$\O' \supset \O \sim \{y \in \Rbf^d, |y| < 1 \}$
and a subgroup $T$ of $\G(\O')$ acting in $\O$ as translations,
$\tau(a)$, sending $y$ to $y +a $, for $y , y+a \in \O$. 
Given $x_0 \in \O$, for all $x$ in $\O$
there is a (unique) translation, $\tau(x-x_0)$ sending  $x_0$ to $x$.
For all $g \in \G(\O)$, two applications of eq.\,(\ref{cocycle}) give  
\be \label{VVV} 
V^{-1}_{\tau (gx -x_0)} (x_0) \, V_{g}(x) \, V_{\tau (x - x_0)} (x_0)  =
V_{\tau (x_0 - gx) \, g \, \tau (x - x_0)} (x_0) ,
\ee
a.e. in $x, x_0$.
The operators in the r.h.s. are indexed by elements of 
$\G(\O, x_0)$ and, by eq.\,(\ref{cocycle}), give a unitary 
representation of it. 
By the equivalence criterium, eq.\,(\ref{UE}), taking $S(x) = V^{-1}_{\tau (x - x_0)} (x_0)$,
the representation of $\Pi(\O)$ is unitarily
equivalent to that given by
\be \label{VxO}
V^{T, x_0}_g(x)   \eqq
R^{x_0} ( \tau (x_0 - gx) \, g \, \tau (x - x_o) ) \, ,
\ee
with
\be
\label{Rx} R^{x_0}(h)\equiv V_h(x_0)
\ee
a unitary representation of $\G(\O,x_0)$ in $K(\O)$.
Clearly, $V^{T, x_0}_h(x_0) = R^{x_0}(h)$.
For $x = x_\infty$, by the cocycle equation and the invariance of $x_\infty$,
\be \label{Rinf}
R_\infty (g) \equiv V_g(x_\infty) 
\ee
gives a unitary representation of $ \G(\O)$ in $K_\infty(\O)$. 
For different $x_0 \in \O$ the groups
$\G(\O,x_0)$ are isomorphic and their
representations $R^{x_0}$ are unitarily equivalent.

\noindent
Conversely, for given $x_0$,  $T$, by eqs.\,(\ref{VxO})(\ref{Rinf}), any pair of (strongly continuous)
representations $R, R_\infty$ of
$\G(\O,x_0)$, $\G(\O)$ in separable Hilbert spaces $K$, $K_\infty$ define 
operators $V^{T, x_0}_g(x)$ satisfying eqs.\,(\ref{cocycle}) and, 
 by Proposition 2.2, 
a cocycle-regular representation $\pi_{R, R_\infty}$ of $\Pi(\O)$.
By eqs.(\ref{VVV}),(\ref{VxO}), the representations of
$\Pi(\O)$ given by  $R$, $R_\infty$, 
are unitarily equivalent for different $T, x_0$.

\vspace{2mm}

Examples of  non-trivial $R$ are given by representations $\rho_n$ of $\Pi(\O)$ in
$L^2 (\O \times \O \times \ldots \times \O, \, d\mu(x_1)  \ldots d\mu(x_n))
\equiv L^2 (\O^n) $
defined by
$$
\rho_n(\alpha(x))  \psi(x_1, \ldots , x_n) = \alpha(x_1) \, \psi(x_1, \ldots , x_n) \, ,$$
$$
\ \ \rho_n(U(g)) \psi(x_1, \ldots . x_n) \, ,
= \psi(g^{-1}x_1, \ldots , g^{-1}x_n) \,
\Pi_i J_g (x_i)^{1/2}    \, .$$ 
Here, $K = L^2 (\O^{n-1})$ and
$\G(\O,x)$ is represented there by the unitary change of variables
$\Pi_{i=2}^n C_g(x_i) $. In these examples, $K_\infty = 0$.

\subsection{Locally Schroedinger representations}

The representations of $\Pi(\O)$ of the above examples describe
states of $n $ particles in $\O$,
  the position of the first particle being described by $\alpha(x)$ and the other
particles  by the variables $U(g)$.
Thus,
\emph {already for local algebras and independently of the topology of $\M$},
 additional conditions are needed
in order to select the representations  of $\Pi(\M)$
with a local one-particle interpretation.

Given a cocycle-regular representation $\pi$ of $\Pi(\M)$,
the unitary equivalence class of the representations $R(\O,x)$, $R_\infty(\O)$
do not depend on $\O$, by the argument before Proposition 2.4.
A non-trivial $ R(\O,x)$ describes
an additional action of $U(g)$,  $g \in \G(\O)$, 
besides the change of variables $C_g$, i.e., 
\emph {additional localized degrees of freedom} with respect to the
Schroedinger positions and momenta in $\O$.
Similarly, a non trivial $R_\infty(\O)$ describes, in $K_\infty(\O)$,
an action of $\G(\O)$ on states localized outside $\O$ and must be excluded
if $\pi(\G(\O))$ has to act locally.

The same conclusion is obtained considering that the 
regions $\O$ are diffeomorphic to $\Rbf^d$, so that the representations
of $\Pi(\O)$ should be compared (they cannot be distinguished in a diffeomorphism
invariant way) with those arising in a diffeomorphism invariant
formulation of QM of a particle in $\Rbf^d$.

In fact, the condition $R(\Rbf^d,x) =  R_\infty(\Rbf^d) = I$ identifies
the Schroedinger representation 
$\pi_S(\Pi(\Rbf^d))$, up to unitary equivalence, multiplicity and the
addition of a trivial representation (``states at infinity'');
$R(\O,x) = R_\infty(\O) = I$
identifies the Schroedinger representation $\pi_S(\Pi(\O))$ 
in $ L^2(\O, d\mu)$, apart from unitary equivalence, 
multiplicity and the addition of a trivial representation,
given by $\pi (\alpha(x)) = 0$, $\pi (U(g)) = 1$,
always allowed for the description of states localized in
$\M \setminus \O$.

Thus, we are  led to consider representations of $\Pi(\M)$
with $\ R(\O,x) = I$ and $R_\infty(\O) = I$, for all $\O \subset \M$.
Since $R_\infty(\O) = I$ also implies
$R_\infty = I$, i.e., a trivial representation of $\Pi(\M)$ in $K_\infty$,
we set in the following $K_\infty = 0$ and therefore, by Proposition 2.2, consider
representations of $\Pi(\M)$ in $ L^2(\M, d\mu) \times K$.

By the above analysis, such conditions \emph {fix the representations
of the local algebras}: by Propositions 2.2, 2.4,
each of them is unitarily equivalent, in $ L^2(\M, d\mu) \times K$,
to a representation
with $V_g(x) = I$ for all $g \in \G(\O)$, i.e.,
to the restriction of $\pi_S(\Pi(\M))$ to $\Pi(\O)$,
hereafter called
\emph {the Schroedinger representation of $\Pi(\O)$ in $ L^2(\M, d\mu)$},
with a multiplicity given by $K$.

\begin{Definition}
A representation $\pi$ of $\, \Pi(\M)$, with $K_\infty = 0$,
is {\bf \em { Locally Schroedinger (LS)}} if it is cocycle-regular and,
for all $\O \subset \M$,
$R(\O,x) $, $R_\infty(\O)$ are the identity;
equivalently, if the representations of $\Pi(\O)$ are unitarily equivalent
to their Schroedinger representation in $ L^2(\M, d\mu)$,
apart from a (common, at most denumerable) multiplicity.
\end{Definition}

The LS condition is a non-trivial restriction already at the algebraic level,
since $\pi_S(\Pi(\O))$ is not a faithful representation of $\Pi(\O)$.
In fact, e.g., for $U(g), \, \alpha \in \Pi(\O) $,
if $g x = x$, $\forall x \in $ Supp $\alpha(x)$,
$ \pi_S ((U(g) - 1) \alpha = 0) $ 
whereas, in $\Pi(\O)$,
$|| (U(g) - 1) \alpha || = 2 \,  \mbox{\rm Sup} |\alpha(x)| $ .

\vspace{2mm}

\noindent
{\bf Remark.}
The LS condition is also
equivalent to the local validity of the Lie-Rinehart relations 
of the generators of $\difm$ as a module on $\cmo$ \cite{noi},
allowing to express them in terms of $d$ independent momenta:
\be \label{LR}
{T_{\sum_i \a_i v_i} = \ume \sum_i (\a_i
T_{v_i} + T_{v_i} \a_i), \,\,\,\,\,\forall \a_i \in  C^\infty_0(\O),\,\,
\,\forall\, v_i \in \L(\O)\,.}\ee
Eqs.(\ref{LR}) hold in fact in the Schroedinger representation and the converse
follows from the proof of Theorem 3.5 in \cite{noi}.

\vspace{2mm}
By the LS condition, for any region $\O$,
one may take, as the observable algebra in $\O$,
the Von Neumann closure $\A(\O)$ of $\pi_S (\Pi(\O))$.
As operator algebras in the space of the
sum, $\rho(\Pi(\M))$, of the LS representations of $\Pi(\M)$,
the Von Neumann algebras $\A(\O)$ generate a $C^*$ algebra 
$\A(\M) $, which can be taken as the (one-particle)
{\em observable algebra} associated to the entire manifold.
$\A(\M)$ contains $\rho(\Pi(\M))$ as a weakly dense subalgebra, so that
representations of $\A(\M)$ define representations of $\Pi(\M)$ and
unitarily equivalent LS representations of $\Pi(\M)$ define
unitarily equivalent representations of $\A(\M)$.
Therefore, the LS representations of $\Pi(\M)$ coincide with the 
\emph {locally normal} (\cite{Haag}, p.131) representations of $\A(\M)$.

By irreducibility of the Schroedinger representation of $\Pi(\O)$
in $L^2(\O, d\mu)$,
the algebras $\A(\O)$ are isomorphic to the algebra $\B(\H)$ of all
bounded operators in a separable Hilbert space. However,
the Schroedinger condition is purely local and the
immersion of the local algebras in $\A(\M)$ gives rise to a non trivial ``bundle''
structure, associated to the topology of $\M$, which 
plays the essential role in the classification of
the representations of $\A(\M)$.

\subsection{Classification of the LS representations by the fundamental group}

By Proposition 2.2 and Definition 2.5, a LS representation of 
$\Pi(\M)$ is given by a collection of representations
of the local algebras $\Pi(\O), \O\subset \M$, all in 
$L^2(\M, d\mu) \times K$ and 
determined by unitary local intertwiners with the Schroedinger representation.
The associated cocycles are locally trivial and our aim is to characterize
the cocycles $V_g(x)$, $g \in \G_L(\M)$, (eq.(\ref{Ug})) associated to \emph {products} of localized
diffeomorphisms; by Proposition 2.2, they identify the representation of  $\Pi(\M)$. 
          The analysis will consist of the following steps.
\vspace{2mm}

\noindent
a) \emph {Local Schroedinger intertwiners}

\smallskip
The following Proposition characterizes the local cocycles for a LS representation of $\Pi(\M)$.

\begin{Proposition}
A regular representation $\pi$ of $\Pi(\M)$, 
with $K_\infty = 0$,
is LS iff for (one and therefore) all $ \O$,
there exist (unique) weakly measurable unitary operators in $K$, 
$W_\O(y,x)$, $x, y \in \O$ 
such that, $\forall g \in \G(\O)$,
\be
\label {VlocS}  V_g(x) = W_O(gx,x) \, , \ \  \forall x \in \O \, ;
\ \ V_g(x) = 1, \  \forall x \notin \O \, . 
\ee
They satisfy
 \be  \label {WW}
W_\O(y,x) W_\O(x,z) = W_\O(y,z) \ , \ \ \ \ \ \ W_\O(y,x) = W_\O(x,y)^{-1} \, ,
\ee
\be \label{indipO}
W_{\O_1} (y,x) = W_{\O_2} (y,x) \, , \
\forall x,y \in \O_1 \, , \ \ \O_1 \subset \O_2. \, 
\ee
\end{Proposition}

\noindent
\Pf
By eq.\,(\ref{UE}) the quasi-equivalence of the representation  
of $\Pi(\O)$ to its Schroedinger representation in $L^2(\M, d \mu)\times K $
implies the existence of  weakly measurable unitary operators $W_\O(x)$ in $K$,
such that, $\forall g \in \G(\O)$, 
\be
\label {VlocS0}   
  V_g(x) =  W_\O(gx) \, W_\O(x)^{-1}   \, ,  \ \ \forall x \in \O \, , \ \ \
V_g(x) = 1 \ , \ \ \forall x \notin \O \, . 
\ee
Then,
\be \label{defW}
W_\O(y,x) \equiv W_\O(y) \, W_\O(x)^{-1}   
\ee
is well defined for all $x,y \in \O$ and weakly measurable, since $W_\O(x)$
is measurable in $x$.
Eqs.(\ref{VlocS}), (\ref{WW}), (\ref{indipO})
immediately follow from the definitions.

\noindent
Conversely, eqs.(\ref{VlocS}),(\ref{WW}) imply
eqs.(\ref{VlocS0}), with $W_\O (x) \equiv W_\O(x,x_0)$
(for any choice of $x_0 \in \O$, apart from a zero measure subset).   
Therefore, the representation of $\Pi(\O)$,
in $L^2(\M, d \mu)\times K $,
is unitarily equivalent to the Schroedinger
representation in the same space, with multiplicity given by $K$.\goodbreak

\vspace{2mm}

\noindent
b) \emph {Homotopy classes of products  of local intertwiners}

\smallskip
As a next step, on the basis of eqs.\,(\ref{VlocS}) (\ref{WW}) (\ref{indipO}), 
we show that, for $g = g_n \ldots g_1, \, g_i\in \G(\O_i)$,
$V_g(x) $ is given by a string of local factors
$W_{\O_i}(x_{i+1}, x_i)$, $x_{i+1} = g_i x_i$;
the resulting operators will be shown to be
invariant under small displacements of the intermediate points $x_i$
and therefore indexed by the homotopy class of the corresponding path.

In fact, given $g \in \G_L(\M)$, $g = h_N \ldots h_1$,
by eq.(\ref{cocycle})  
$V_g(x)$ is represented by  a product of terms of the form $V_{h_i} (y_i)$.  
Dropping all the factors with $h_i y_i = y_i$
(equal to the identity by eq.(\ref{VlocS})),
we are left with a subset $g_1 \ldots g_n$, $g_i \in \G(\O_i)$, such that
$$ gx = g_n \ldots g_1 x \ ,  \ \   x_{i} , x_{i-1} \in \O_{i} \,
\ \ x_i \equiv g_i x_{i-1} \neq x_{i-1} \ , \ \  x_0 \equiv x\, , \ x_n \equiv gx.  $$
Then, by eq.(\ref{VlocS}), one has
\be \label{string}
V_g(x) = V_{g_n} (x_{n-1}) \dots V_{g_1}(x) =
W_{\O_n}(gx, x_{n-1})  W_{\O_{n-1}}(x_{n-1}, x_{n-2})
\ldots W_{\O_1}(x_1,x).  \ee

\begin{Lemma}  
Given $x,y \in \M $, a regular path $\gamma(y,x)$ starting at $x$
and ending at $y$, a partition of $\gamma$,
$\gamma = \gamma_n (y,x_{n-1}) \circ \ldots \circ \gamma_1(x_1,x)$,
open sets $\O_1 \ldots \O_n$ with $\O_i \supset \gamma_i$
and unitary operators $W_\O(y, x)$ in $K$ satisfying eqs.(\ref{WW}),(\ref{indipO}),
the operators
\be \label{gamma}
W(y,x,\gamma) \equiv
W_{\O_n}(y, x_{n-1}) \, W_{\O_{n-1}}(x_{n-1}, x_{n-2})
\ldots W_{\O_1}(x_1,x), \ \ \
\ee
only depend on $x, y$ and on the homotopy class $[\gamma ]$ of
$\gamma (y,x)$.
They satisfy the composition law
\be \label{gammadelta}
W(y,x,[\gamma]) \,W(x,z,[\delta]) = W(y,z,[\gamma] \circ [\delta]). 
\ee
For almost all $x$, the {\bf \em topological operators}
$ W(x,x,[\gamma]) $ provide a unitary representation $\R^x([\gamma])$ of
$\pi_1(\M)$, all belonging to the same {\bf \em equivalence class} $[\R]$.

\noindent
Two systems of (weakly measurable) unitary operators, 
$W_1$, $W_2(y,x,[\gamma])$,
satisfying eq.(\ref{gammadelta}),
are related by (weakly measurable) unitaries $S(z)$,
\be \label{equivaperti}
W_2(y,x,[\gamma]) = S(y) \, W_1(y,x,[\gamma]) \, S(x)^{-1} 
\ee 
iff, for some (and then for all) $x \in \M$,
the corresponding systems for closed paths are unitarily equivalent:
\be \label{equivchiusi}
\R_2^x([\gamma ]) = S(x) \, \R_1^x([\gamma ]) \, S^{-1}(x) \, ,
\ee
equivalently, iff the corresponding equivalence classes
$[\R_i]$, $i = 1, 2$, coincide.
\end{Lemma}

\noindent
\Pf Given two choices $\{\gamma_i, \,\O_i\}$ and $\{\gamma'_i, \,\O'_i\}$,
a ``combined'' partition of $\gamma$ is defined by
the sequence $\{y_k\}$ obtained by ordering the points
$x_i, x'_j$; each piece $\gamma''_i(y_i, y_{i-1})$ of the so obtained partition is
contained in some $\O''_i \subset \O_k \cap \O'_{k'}$ for some pair $k, k'$.
Therefore, by eq.(\ref{indipO}) the operators $W(y,x,\gamma)$ associated to
$\{ \gamma''_i, \,O''_i\} $
coincide with both the operators constructed from
$\gamma_i$ and $\gamma'_i$.
Since the sets $\O_i$ can be kept fixed for a small deformation of $\gamma$,
the result only depends on the homotopy class of $\gamma$. 
Independence from the partition of $\gamma \circ \delta$
immediately implies eq.(\ref{gammadelta}).

\noindent
By  eqs.(\ref{gammadelta}), for all $x,y$, fixed a path $\delta$ from $y$ to $x$,
$$ [\gamma] \to [\delta^{-1} \circ \gamma \circ \delta] $$
is a bijection between (the equivalence classes of) the closed paths
with base point $x$ and those with base point $y$, and
\begin{equation} \label{Wyx} 
W(y,y,[\delta^{-1} \circ \gamma \circ \delta]) =
W(x,y,[\delta])^{-1} \, W(x,x,[\gamma]) \, W(x,y,[\delta]).  
\end{equation}
Then, eq.(\ref{equivaperti}) implies
eq.(\ref{equivchiusi}) yielding 
the unitary equivalence of the
representations $\R^x$ of $\pi_1(\M)$ for different $x \in \M$; conversly, 
it is easy to check that,
given $x$ and $S(x)$ satisfying eq.(\ref{equivchiusi}),
\be
S'(y)  \equiv W_2 (y, x, \gamma) \, S(x) \, W_1 (y, x, \gamma)^{-1}  
\ee
satisfies eq.(\ref{equivaperti}), for any choice of curves $\gamma$
starting at $x$ and ending at $y$ and it is measurable in $x$
if the curves depend continuously on $x$ in an open set $\M_0$ with
complement of zero measure, as in the proof of Proposition 2.4.

\vspace{2mm}

\noindent 
c) \emph {Classification by the fundamental group}

\smallskip
By the above results, in a LS representation $\pi$ of $\Pi(\M)$,
the operators $\pi(U(g))$ representing a product
of localized diffeomorphisms, $g = g_n \ldots g_1$, are given by
eqs.(\ref{Ug}), with
\be \label{rapprV}
 V_g(x) =  W(gx,x,[\gamma_g]) \, ,
\ee
where $W(gx,x,[\gamma_g])$ is defined by eqs.(\ref{string}),(\ref{gamma}) and $\gamma_g$ is 
the integral curve, from $x$ to $gx$, defined by $g_n \dots g_1$.

By eq.(\ref{UE}) two LS representations $\pi_i$ of $\Pi(\M)$ are unitarily equivalent
iff the corresponding  $W$ operators are related by eq.(\ref{equivaperti}),
and therefore, by Lemma 2.7, iff the associated equivalence classes
$[\R_{\pi_i}]$ of representations of $\pi_1(\M)$ coincide.
Hence, the LS representations of $\Pi(\M)$ are classified
by the corresponding unitary representations of $\pi_1(\M)$.
The following Lemma shows that \emph{all} the unitary representation of
$\pi_1(\M)$ appear in the classification.

\begin{Lemma}
Any unitary representation $\R$ of
$\pi_1(\M)$  in a (separable) Hilbert space $K$ 
defines a LS representation
$\pi_\R$ of $\Pi(\M)$ in $L^2(\M, d \mu)\times K $, 
with the same equivalence class,  $ [\R_{\pi_\R}] = [\R] $ 
\end{Lemma}

\noindent
\Pf
By Proposition 2.2, given $\R$ we have to construct 
weakly measurable unitary operators $V_g(x)$ satisfying eq.(\ref{cocycle}).

\noindent
To this purpose, we fix a point $x_0 \in \M$ and associate to each
$ x \in \M$ a path $\delta(x, x_0)$ from $x_0$ to $x$.
As in the proof of Proposition 2.4, 
such paths can be taken continuous in $x$, in the $C^0$
topology of paths, for all $x$ in a set $\M_0 \subset \M$, with a complement of
zero measure.
Then, to each
$\gamma(y,x)$ from $x$ to $y $, $\ \forall x,y \in \M$, we associate
the closed path
\be \label {betadigamma}
\beta(\gamma(y,x)) \equiv \delta(x_0, y) \circ \gamma(y,x) \circ \delta(x, x_0) \, ,
\ee
with $\delta(x_0, y) \equiv \delta(y, x_0)^{-1}$.
Clearly, the equivalence class $[\beta]$ only depends on $[\gamma]$.
Composing two paths, $\gamma(y, x)$, $ \gamma(x, z)$,
gives rise to the composition of the corresponding images in $\pi_1(\M)$:
$$
\beta(\gamma(y,x) \circ \gamma(x,z)) = 
\delta(x_0, y) \circ \gamma(y,x) \circ \delta(x, x_0)
\circ \delta(x_0, x) \circ \gamma(x,z) \circ \delta(z, x_0)
= $$
\be \label{betagruppo} = \beta (\gamma(y,x)) \circ  \beta (\gamma(x,z)). \ee

\noindent
Given a unitary representation $\R$ of $\pi_1(\M)$ in $K$,
we then define, for all $g \in \G_L(\M)$, $x \in \M$ (omitting for simplicity the
equivalence class notation for the paths)
\be  \label{defvg}
V_g(x) =  W(gx,x,\gamma_g) \equiv \R (\beta (\gamma_g(gx,x))),  
\ee
with, as above, $\gamma_g(gx,x)$ the  integral curve
associated to $x$ and $g$. 
For $gx =  x = x_0$, eq.(\ref{defvg}),
identifies $\R([\gamma_g])$ with the topological factors $W(x_0, x_0, \gamma_g)$
representing $\pi_1(\M)$ with base point $x_0$. 

\noindent
By the above continuity property of paths, 
$V_g(x)$ is locally constant in $x$ and $gx$ in $\M_0$; it
is therefore strongly continuous in the parameters of one-dimensional 
subgrous of $\G_L(\M)$ and defines a cocycle:
by eqs.(\ref{defvg}),(\ref{betagruppo}), $\forall x \in \M$
$$ \vg(h x) \vh(x) =
\R(\beta(\gamma_g(ghx, hx))) \, \R(\beta(\gamma_h(hx, x))) 
= $$
$$ = \R(\beta(\gamma_{gh} (ghx, x))) = V_{gh} (x) \, .
$$
We have therefore a unitary representation of $\Pi(\M)$
in $L^2(\M, d \mu)\times K $, which is regular and cocycle-regular,
with $ \R $ as the associated representation of $\pi_1(\M)$. 

\noindent 
For the proof of the  LS property, eqs.(\ref{VlocS}),(\ref{VlocS0}), 
consider $x, z \in \O $, $g \in \G(\O)$; since $[\beta]$ only depends on
$[\gamma]$, using  eq.(\ref{betagruppo}), $\forall \gamma(g x, z),
\,\gamma(z,  x) \subset \O, $
$$
  V_g(x) = \R(\beta([\gamma_g(gx, x)])) =
  \R(\beta([\gamma(gx, z)])) \,
  \R(\beta([\gamma(z, x)])) \, ,
$$
which is of the form of eq.(\ref{VlocS0}) since 
$ \R(\beta([\gamma(y, z)]))$ only depend on $y$, for fixed $z$.

\vspace{1mm}

Eq.(\ref{defvg}) also provides a direct relation, in $\pi_\R$, between the
topological operators associated to closed paths with different base points:
by using the existence, for any closed $\gamma$, 
of a path $\gamma_g$ with the same base point $x$ and $[\gamma_g] = [\gamma]$,
one has
\begin{equation} \label{Wxx0}
W(x,x,\gamma) =  \R(\beta([\gamma])) = W(x_0,x_0,\beta(\gamma)).   
\end{equation}
In conclusion, we have:
\begin{Theorem} 
The  Locally Schroedinger representations $\pi$ of $\Pi(\M)$, equivalently,
the locally normal representations of the observable algebra $\A(\M)$,
are given by eqs.(\ref{Ug}),(\ref{rapprV}).
Up to unitary equivalence, they are classified
by the associated unitary equivalence class of
representations of $\pi_1(\M)$, $[\R]$.
Any unitary representation $\R$ of $\pi_1(\M)$ (in a separable space)
define a LS representation $\pi_\R$ of $\Pi(\M)$, equivalently, 
a locally normal representations of $\A(\M)$, with $[\R]$ the associated
representation of $\pi_1(\M)$.
\end{Theorem}

\begin{Corollary}
The commutant and the centre of $\pi_\R (\A(\M))''$,
in the representation space $L^2(\M, d \mu) \times K$,
are given by the commutant $\R'$ and
the centre $\R' \cap \R''$ of the representation $\R$ of $\pi_1(\M)$ in $K$:
\begin{equation}
\pi_\R(\A(\M))' = I \times \R',
\end{equation}
\begin{equation} \label{centre}
  \pi_\R(\A(\M))' \cap \pi_\R(\A(\M))'' = I \times (\R' \cap \R'').
\end{equation}
\end{Corollary}

\begin{Corollary} (Analogue of Von Neumann uniqueness theorem)
If $M$ is simply connected, the locally normal representation of $\A(\M)$ is unique, up to multiplicity and
coincides, up to unitary equivalence, with the Schroedinger representation in $L^2(\M, d \mu)$.
\end{Corollary}

\noindent
{\em Proof of Corollary 2.10}. A (bounded) operator $A$ in $L^2(\M, d \mu)\times K $ commuting with
all the multiplication operator $\pi_\R(\alpha(x))$ is a multiplication operator
$A(x)$ in $K$.
By eq.(\ref{betadigamma}) with $x = x_0$, for all $y \in \M$,
$\beta (\delta (y, x_0)) $ is the identity. Taking
a diffeomorphism $g$, sending $x_0$ into $y$,
such that $ [\gamma_g(y, x_0)] = [\delta (y, x_0)]$, 
the vanishing of the commutator of
$A(x)$ with all the operators $\pi_\R(U(g))$ implies, by eq.(\ref{defvg}),
that $A(x) $ is constant, a.e. in $x$, i.e., $A = I \times A_K $.

\noindent
Since $A$ commutes with $\pi_\R(U(g))$, $A_K$ commutes with $V_g(x)$ and  
therefore, by eq.(\ref{defvg}), it commutes with the representation $\R$ of
$\pi_1(\M)$ in $K$. Conversely, by eq.(\ref{defvg}), operators $I \times B$,
with $B \in \R'$, commute with $\pi_\R(\A(\M))$.
This immediately implies
$$
\pi_\R(\A(\M))'' = \B (L^2(\M, d\mu)) \times \R''
$$
and eq.(\ref{centre}) follows.

\subsection{Topological observables}

Theorem 2.9 and Corollary 2.10 classify the representations of $\A(\M))$
in terms of weak limits of local observables, given by the centre of the
associated representation $\R$ of $\pi_1(\M)$.
They leave open the question of
whether the operators of $\R$, and those of $\R' \cap \R''$,
already belong to the ``quasi local'' observable algebra
$\A(\M)$.

\begin{Theorem}

\noindent  
 For all $x \in \M$
there exists $\O \ni x$ such that, for all $\R$,  $\pi_\R(\A(\M))$ contains 
the topological operators $ W( x, x, \gamma) \, P_\O $,
eq.(\ref{gamma}), with $P_\O$ the projection on $L^2(\O, d\mu) \times K$;
actually they belong  
 to (the representation of) the
algebra generated by a finite number of algebras
$\A(\O_i)$, $\O_i$ contained in a neighbourhood of $\gamma$.
 
\noindent
Furthermore, if   $\M$ is compact

\noindent
i) $\A(\M)$ contains a ``topological'' $C^*$-subalgebra $\T(\M)$,
isomorphic to the group algebra of $\pi_1(\M)$
(generated in norm by the sum of its unitary representations).
For all $\R$, $\pi_R(\T(\M))$ is generated by the operators
$ W( x_0, x_0, \gamma)$ representing $\pi_1(\M)$ in $K$ and
the representations of $\T(\M)$ classify the (locally normal)
representations of $\A(\M)$.

\noindent
ii) if $\pi_1(\M)$ is finite,
$\T(\M)$ is a Von Neumann algebra and its centre $\cal {Z}$
classifies the (locally normal) representations of
$\A(\M)$;

\noindent
iii) if $\pi_1(\M)$ abelian, the spectrum of $\T(\M)$
labels the factorial (locally normal) representations of
$\A(\M)$.
\end{Theorem}

\noindent

\Pf The topological factors $W(x,x, [\gamma])$ can be represented
in terms of localized observables as follows.
Given  $g = g_n \ldots g_1$, $g_i$ localized in $\O_i$, 
with  $ g x =  x$ for some $ x \in \O_1$,
by eqs.\,(\ref{rapprV}),(\ref{Ug}) one has, $\forall y \in \O_1$
\be \label{Wgamma}
W(gy,y, [\gamma_g]) = V_g(y) = C_g^{-1}  \pi(U(g_n) \ldots U(g_1)) \, ,
\ee
with $\gamma_g(gy,y) \subset \cup_i \O_i \, $; moreover,
for all $y \in \O \subset \O_1$, such that $g (\O) \subset \O_1$,
by eqs.\,(\ref{gamma}),(\ref{WW}), one has 
\be \label {splitgamma}
W(gy,y, [\gamma_g]) = W_{O_1}(gy, x) \, W( x,  x, [\gamma_g(x, x)]) \,
     W_{O_1}(x, y). 
\ee
In a representation $\pi_\R$,
%for all $x \in \M_0$ (see the proof of Lemma 2.8)
one can choose $\O_1$ such that, by eq.(\ref{defvg}), 
the operators $W_{\O_1}$ in eq.(\ref{splitgamma}) are the identity,
and therefore
\be \label{OTF}
W( x, x, [\gamma_g]) \, P_\O =
C_g^{-1} \, \pi_R ( U(g_n) \ldots U(g_1)) \, P_\O \, ,
\ee
with $P_\O$ the projection on $L^2(\O, d\mu) \times K$.

\noindent $ \pi_\R( U(g_n) \ldots U(g_1)) \, P_\O $ maps
$L^2(\O, d\mu) \times K$ into
$L^2(g(\O), d\mu) \times K$.
A projector $P_{g(\O)}$
can therefore be inserted in the r.h.s. of eq.\,(\ref{OTF}) and
$C_g^{-1} \, P_{g(\O)}$,
as an operator in $L^2(\O_1, d \mu)$, belongs to the (weakly closed)
algebra $\A(\O_1)$, by irreducibility of the Schroedinger
representation of $\A(\O_1)$ in $L^2(\O_1, d \mu)$.
                                   
\noindent
 By the argument before eq.(\ref{Wxx0}), eq.(\ref{OTF}) applies
to all $[\gamma] \in \pi_1(\M)$,
with almost any $x$ as base point, and therefore gives 
the unitary representations of $\pi_1(\M)$, 
characterizing the representations of $\A(\M)$ directly in terms of
observables localized along the paths $\gamma$.

\noindent
i) Since $ \gamma \to \beta(\gamma)$ has an inverse
$$
\gamma(\beta) = \delta(x, x_0) \, \beta \, \delta(x_0,x) \, ,
$$
by eq.(\ref{Wxx0}) one has
\begin{equation} \label{Rbeta}
\R(\beta) = W(x,x,\gamma(\beta))   
\end{equation}
with $\beta$ any closed path with base point $x_0$ and
$x$ almost any point in $\M$.

\noindent
Since, as above, there is a
$\gamma_g$ in the equivalence class of
$\gamma(\beta)$, eqs.(\ref{Rbeta}),(\ref{OTF}) imply
\begin{equation} \label{RbetaPO}
  \R(\beta) P_\O = W( x, x, [\gamma_g]) \, P_\O 
\end{equation}  
for all closed paths $\beta$ with base point $x_0$, almost all $x \in \M$,
$\gamma_g \in [\gamma(\beta)]$.

\noindent 
If $\M$ is compact, it may be covered by a finity family $\{\O_\alpha\}$ and therefore
by disjoint measurable sets $O_\alpha \subset \O_\alpha$; multiplying eq.(\ref{RbetaPO})
on the right by $P_{O_\alpha}$ and summing over $\alpha$, one gets
\begin{equation} \label{RbetaOss}
  \R(\beta) = \sum_\alpha W( x_\alpha, x_\alpha, [\gamma_{g_\alpha}]) \, P_{O_\alpha}
  \equiv \pi_\R (A (\{ x_\alpha , g_\alpha \}))
\end{equation}
with $ A (\{ x_\alpha , g_\alpha \}) \in \A(\M)$, by eq.(\ref{OTF}).

\noindent
Since this holds for any $\R$ and all LS representations of $\Pi(\M)$
are unitarily equivalent to a $\pi_\R$, the observable algebra 
$\A(\M)$, which has been defined in the sum
of the LS representations of $\Pi(\M)$, contains a $C^*$-subalgebra
$\T(\M)$
isomorphic to the group algebra of $\pi_1(\M)$.
By eq.(\ref{Wxx0}), the generators of $\T(\M)$ are represented in
$\pi_\R$ by $W(x_0,x_0,[\beta])$.
By Theorem 2.9, the representations of $\T(\M)$
classify the (locally normal) representations of $\A(\M)$.

\noindent
ii) If $\pi_1(\M)$ is
finite, its representations are finite dimensional,
apart from multiplicities, and therefore norm and weak closures
coincide. 

\noindent
iii) If $\pi_1(\M)$ is abelian, the Von Neumann algebra generated by
the representations of $\T(\M)$ is the algebra of Borel functions
(with the weak topology defined by Borel measures)
on its spectrum, which therefore indexes
the factorial representations. 

\vspace{2mm}

If the diffeomorphism $\tilde g$ defined by $g = g_n \ldots g_1$ belongs to the connected
component of the identity of the diffeomorphism group of $\O_1$,
then $C_g = \pi_\R(U(\tilde g))$  and
the representation of the topological factors,  eq.(\ref{OTF}),
only involves the local  algebras $ \Pi(\O_i)$.
If $\M$ is not orientable, this is not in general the case, and 
$C_g$ only belongs to the weakly closed 
algebra $\A(\O_1)$ \cite{Gian}.

\vspace{2mm}
The classification in terms of a locally Schroedinger description
and unitary representations of $\pi_1(\M)$ 
was also obtained in \cite{noi}, for the crossed product algebra
$C(\M) \times \difm^c$, $\difm^c$ the universal covering group
of Diff$(\M)$, under a ``Lie-Reinhardt'' condition, which, for the local algebras,
is equivalent to the LS property. However,
the starting point is different in the two approaches:
$C(\M) \times \difm^c$ includes from the beginning the
operations of transport on closed paths as independent variables.

 On the contrary, in the present approach, 
the observable algebra is generated by \emph {the local algebras} 
$\A(\O)$, associated to topologically trivial regions,
homeomorphic to $\Rbf^d$, and therefore assumed 
to be represented as in ordinary Schroedinger QM, apart from
unitary equivalence and multiplicity.
Topological effects arise from  
\emph {the collection of local Schroedinger Quantum Mechanical descriptions}.
They appear in products of local observables
and are characterized by the representation of operations of
physical trasport along non-trivial closed paths, constructed
as a sequence of strictly localized operations.

 The result for $C(\M) \times \difm^c$ 
actually follows from the present analysis through the
extension of eq.(\ref{rapprV})
to the representatives $U(g_{\lambda v})$ of all the one-parameter
subgroups of $\difm^c$, which follows, as in \cite{noi}, Proposition 4.4.,
from the identification of the corresponding generators.\goodbreak

\section{The observable and the gauge action of $\pi_1(\M)$}

\subsection{The Dirac approach to QM on manifolds}

\vspace{1mm}
In his treatment of $N$ identical particles, Dirac proposed a solution of two problems:
the explicit identification of the observables and the classification of their representations.
In terms of operator algebras, his strategy was:

\noindent
i) to consider the algebra of $N$ distinguishable particles,
which may be taken as the  Weyl algebra for $3N$ degrees of freedom
$\A_W(3N)$, or as the algebra $\B ( L^2(\Rbf^{3 N}))$ of all bounded operators
in $L^2(\Rbf^{3 N})$,
and to define the observable algebra $\A_S(N)$ of $N$ identical particles
as the subalgebra invariant under the permutation group $S_N$, which thus plays the
role of a gauge group.

\noindent
ii) to obtain representations of $\A_S(N)$ by the
decomposition of the unique (by the Von Neumann theorem) representation of $\A_W(N)$,
or of the unique normal (irreducible) representation of $\B (L^2(\Rbf^{3 N}))$.
Both are defined in $L^2(\Rbf^{3 N})$ and the decomposition is given by their commutant,
generated by the gauge representation of $S_N$ in $L^2(\Rbf^{3 N})$.  
 
\vspace{1mm}
Since the configuration manifold $\M_S$ of $N$ identical particles in three dimensions
(introduced above) has $\S_N$ as fundamental group and $\Rbf^{3 N}$ as universal covering manifold, 
Dirac strategy may be translated as a recepee for Quantum Mechanics on a manifold $\M$,
consisting in the following steps:

\noindent
i) the configuration manifold $\M$ is replaced by its universal cover,
$\tilde{\M}$;

\noindent
ii) the observable algebra $\A(\M)$ is identified as the subalgebra of
$\B(L^2(\tilde{\M}, d \tilde{\mu}))$ ($d \tilde{\mu}$ locally in the equivalence
class of the Lebesgue measure) invariant under the action of
$\pi_1(\M)$ in $L^2(\tilde{\M})$;

\noindent
iii) the  representations of $\A(\M)$ are obtained  by the decomposition
of $ L^2(\tilde{\M})$ according to the representations of $\pi_1(\M)$.

Such an approach had a substantial influence on the treatments of QM on manifolds;
it provides a natural framework for the introduction of $\tilde{\M}$,
also suggested by the Aharonov-Bohm phenomenon and naturally associated 
to the Functional Integral approach, and for the associated role  
of $\pi_1(\M)$ as a gauge group.
The question then arises about the physical justification of the choices underlying 
the Dirac strategy and about the completeness of the classification of the
representations of the  observable algebra obtained in that way.
A related question is whether, in that approach,
the role of the fundamental group is only that of gauge transformations.

The aim of this section is to discuss the relation between Dirac
strategy and the treatment of QM on manifolds discussed in Section 2, which
only relies on physically motivated principles.
The essential results are the following:

\vspace{1mm}
\noindent A.
If $\pi_1(\M)$ is {\bf \em amenable} (i.e.,
it admits an invariant mean, which is always the case for finite and abelian groups), then:
 
\noindent
i) a representations of $\pi_1(\M)$ \emph {by observable operators}
is always present in $L^2(\tilde \M)$, unitarily equivalent to its representation
as a gauge group;

\noindent
ii) the gauge and the observable topological classification of the
representations of $\A(\M)$ in $L^2(\tilde \M)$ are the same. The
observable and gauge representations of $\pi_1(\M)$ associated to
irreducible representions of $\A(\M)$ coincide, apart from a complex conjugation;

\noindent
iii) the irreducible (locally Schroedinger) representations of $\A(\M)$ are \emph{all}
contained in the reduction of $L^2(\tilde \M)$. 

\vspace{1mm}
\noindent B.
If $\pi_1(\M)$ is not amenable,
results i) and ii) still hold, but the reduction of 
$L^2(\tilde \M)$ \emph {does not} contain all the irreducible
(locally Schroedinger) representations of $\A(\M)$, and in fact
even its ordinary Schroedinger representation is \emph {not}
obtained in the reduction.

\vspace{1mm}
In the derivation of the above result, the central role is played by the following facts:

\noindent
i) the gauge representation of $\pi_1(\M)$ in $L^2(\tilde\M)$ is a multiple of
its \emph {right} regular representation;

\noindent
ii) consequently, corresponding \emph {left} regular representations can be constructed
and are observable;

\noindent
iii) \emph {completeness} of the regular representation, i.e. the presence of
all the irreducible representations in its (possibly integral) decomposition is
\emph {equivalent} \cite{Pier}
to \emph {amenability} of $\pi_1(\M)$.

\vspace{1mm}
Non amenable groups may appear for manifolds describing relevant
physical situations, e.g. in the case of a plane with $n$ holes, $n>1$,
where the fundamental group is freely generated by $n$ elements
and therefore non amenable.
\cite{JD}

For non amenable groups, even if the role of $L^2(\tilde\M)$ and of the gauge group are lost, 
a modification of Dirac strategy still applies: all the irreducible
LS representations of $\A(\M)$ can still be obtained by the action of coordinates
and vector fields of $\M$ on Hilbert spaces of wavefunctions on $\tilde\M$,
\emph {with suitable (non $L^2$) scalar products}.

\vspace{2mm}

\subsection{Equivalence of the Dirac approach for amenable $\pi_1$ }

To prove the above results, we have to confront the ``extended Schroedinger'' 
representation of $\A(\cal \M)$ in $L^2(\tilde \M)$ arising in the Dirac strategy
with the classification given by Theorem 2.9.

\vspace{2mm}

\noindent
a)  {\em The standard representation of $\tilde \M$}
\vspace{1mm}

A standard representation of $\tilde\M$, together with the gauge action of  
$\pi_1(\M))$ on it, is obtained as follows: 

Denoting, as before, by $\gamma(y,x) $  the
continuous paths in $\M$ from $x$ to $y$, given 
$x_0 \in \M$, $\tilde \M$ can be identified as the space of pairs
\be{{\tilde \M } = \{ {\xi} \eqq (x, [\gamma]); \, x \in \M, \,\gamma = \gamma(x, x_0)\} \, , }
\label{Mtilde1}
\ee
with manifold structure given by the basis of open sets
\be{  {\tilde \O}_{(x,[\gamma])} \eqq \{ (y, [\gamma_y]);    y \in \O_x \subset \M, \,[\gamma_y] =
  [\gamma(y, x) \circ \gamma(x, x_0)], \gamma(y,x) \subset \O_x \}  \, ,  }
\ee
indexed by and homeomorphic to the small neighbourhoods $\O_x$ of $x$.  
The resulting space covers $\M$ and is simply connected.

$\pi_1(\M))$ acts on $\tilde \M$ by its right action on the paths $\gamma(x,x_0)$,  
\be
 r(\eta) : (x, [\gamma]) \mapsto  (x, \gamma \circ \eta^{-1}) \, ,
\label{gaugegroupM}
\ee
$\eta \in \pi_1(\M)$ with base point $x_0$, and clearly
\be \label{quotient}
\M = \tilde \M / r(\pi_1(\M)) \, .
\ee
$r(\pi_1(\M)) $ has therefore the role of a gauge group, associated to the redundant description
given by$\tilde \M$. No identification of $\M$ with a subset of $\tilde \M$ is assumed at this point,
nor is it implied by eqs.(\ref{gaugegroupM})(\ref {quotient}). 

\vspace{2mm}
\noindent
b)  {\em The standard representation of $\A(\cal \M)$ in $L^2(\tilde \M)$}
\vspace{1mm}

A standard representation of $\A(\cal \M)$ in $L^2(\tilde \M)$ is provided
by multiplication of pairs  $(x,[\gamma])$ by functions of $x$ and by the action
of localized diffeomorphisms of $\M$ on $\tilde \M$.
In fact, for all $\O$, diffeomorphisms of $\M$ localized in $\O$, $g_{\lambda v}$, define
diffeomorphisms of ${\tilde \M}$, by
$$ g_{\lambda v}(x, [\gamma]) \eqq (g_{\lambda v} x, [\gamma_g(gx, x) \circ \gamma] ), $$
with $\gamma_g(gx,x)$ the integral curve of $v$, as above (the 
homotopy class of the curve in the r.h.s. only depending on $[\gamma]$). 

On $\tilde \M$ we consider measures $d\tilde\mu$, locally in the class of the Lebesgue measure
on the disks to which the ${\tilde \O}_x$ are homeomorphic. 
$\A(\M)$ is then represented in $L^2(\tilde \M, d\tilde\mu)$
by its ``extended Schroedinger'' representation ${\tilde \pi}_S$,
with action given by equations of the form of eqs.(\ref{Ug}),
with $K$ one-dimensional and $V_g = 1$
\be
\tilde\pi_S(\a)  \psi(x,[\gamma]) =  \a(x) \psi(x,[\gamma])  \ \ \ \ \ \label{aMt}
\ee
\be
\tilde\pi_S (U(g_{\lambda v})) \psi(x,[\gamma]) =  \psi (g^{-1} x,
[\gamma_g(g^{-1}x,x) \circ \gamma]) \, J_g(x,[\gamma])^{1/2} \, . \label{UMt}
\ee
As above,
$J_g(\xi) \equiv [d\tilde\mu(g^{-1} \xi)/ d\tilde\mu(\xi)]$, $\xi \equiv (x,[\gamma])$.

\vspace{1mm}
Eqs.(\ref{aMt}),(\ref{UMt}) reproduce the representation adopted in the Dirac approach;
actually, if $\A(\tilde \M)$ is identified with the Dirac ``enlarged algebra'', 
one of the Dirac steps, i.e. the choice of its Schroedinger representation,
is forced by the uniqueness result for simply connected manifolds, Corollary 2.11.

\noindent
The (Dirac) gauge representation of $\pi_1(\M)$ is given by its
right action in $\tilde \M$, which defines a unitary representation 
$$
 \psi (x, [\gamma]) \mapsto \psi (r(\eta^{-1})(x, [\gamma])) \, J^\eta (x,[\gamma])^{1/2}  
$$
in $L^2(\tilde \M, d\tilde\mu)$, commuting with $\tilde\pi_S(\A(M))$
by eqs.(\ref{aMt}),(\ref{UMt}),
with $J^\eta$ a Jacobian factor as in eq.(\ref{UMt}).

\noindent
If $\pi_1(\M)$ is finite, since localized diffeomorphisms of $\M$ act
in finite union of small regions in $\tilde \M$, $\tilde\pi_S (\A(\M))$
is contained in $ \pi_S (\A (\tilde \M))$ and coincides with its 
subalgebra invariant under the gauge representation of $\pi_1(\M)$.
\goodbreak

\noindent 
c) {\em Equivalence of $\tilde\pi_S(\A(\M))$ to a representations in  $L^2(\M) \times l^2(\pi_1(\M))$ }
\vspace{1mm}
\def \tildeM  {${\tilde \M}$}

In order to confront $\tilde\pi_S(\A(\M))$ with the classification in Theorem 2.9,
it is convenient to convert it to the form given by Proposition 2.2.
Introducing $\delta(x, x_0)$ and $\beta(\gamma)$ as in the proof of Lemma 2.8, eq.\,(2.26),
$$\beta(\gamma(x, x_0)) \equiv \delta(x_0, x) \circ \gamma(x, x_0) \, , $$
$[\beta]$ depends only on $[\gamma(x, x_0)]$ and belongs to $\pi_1(\M)$, with $x_0$ as base point.
The correspondence is invertible and $\tilde \M$ can therefore be represented as  
\be{ {\tilde \M } = \{ {\tilde x} = (x, [\beta]); \, x \in \M, \, [\beta]\in \pi_1(\M) \} =
  \M \times \pi_1(\M) \, , }\ee
In this representation, $\tilde \M$ consists of copies of $\M$, indexed by $\pi_1(\M)$ and permuted by
its action, eq.(\ref{gaugegroupM}). Clearly, also the \emph {left} action of $\pi_1(\M)$,  
\be
 l(\eta) : (x, [\beta]) \mapsto  (x, [\eta \circ \beta]) \, ,
\label{obsgroupM}
\ee
acts by permutation of the copies of $\M$ in $\tilde \M$ and \emph {commutes} with the right action.
Contrary to the right action, the left action \emph {depends} on the above construction,
indexed by the family of paths $\delta(x, x_0)$, equivalently, by the corresponding
family of embeddings of $\M$ in $\tilde \M$.

Actually,
as it is clear by a transformation to the first representation
of $\tilde \M$, eq.(\ref{Mtilde1}), 
the above left action of $\pi_1(\M)$
only depends upon the family of isomorphisms between
the fundamental groups with different base points $x \in \M$ given by
the paths $\delta(x, x_0)$. It is immediate to verify that such isomorphisms
are \emph {unique} (i.e. $\delta(x,x_0)$ independent) iff $\pi_1(\M)$ is \emph {abelian};
in this case, the right and left actions coincide, up to an inversion.
                                   
Since, apart from a set of zero measure, all points of $\tilde \M$ have a neighbourhood
with $[\beta]$ fixed, the measure on $\tilde\M$ can be chosen, in the same measure class,
as $d\tilde\mu (x, [\beta]) = d\mu(x)$, independent of $[\beta]$. This provides the
unitary equivalence
\be{
L^2(\tilde \M, d\tilde\mu) \sim L^2(\M, d\mu) \times l^2(\pi_1(\M)). }\ee
In this space, the representations $\tilde\pi_S (\A(\M)$ takes the form,
for simplicity without change in notation, 
\be
\tilde\pi_S (\a)  \psi(x,[\beta]) =  \a(x) \psi(x,[\beta])  \ \ \ \ \ \label{aMt1}
\ee
\be
\tilde\pi_S (U(g_{\lambda v})) \psi(x,[\beta]) =
\psi (g^{-1} x , [\theta^{-1} (g, x) \circ \beta])
J_g(x)^{1/2} \, . \label{UMt1}
\ee
with
\begin{equation}
\theta^{-1}(g, x)  \equiv
\delta(x_0, g^{-1}x) \circ \gamma_g(g^{-1}x,x) \circ \delta(x,x_0) \in \pi_1 (\M) \, .
\label{thetadigamma}
\end{equation}

\vspace{2mm}
\noindent
d) {\em The left and right regular representation of $\pi_1(\M)$ in  $L^2(\tilde\M)$}
\vspace{1mm}

Eq.(\ref{UMt1}) gives the action of $U(g)$ in terms of the 
{\bf \em left regular representation} $\R_l$ of $\pi_1(\M)$ in $l^2(\pi_1(\M))$,
\be
 (I \times \R_l(\theta)) \, \psi (x, [\beta]) = \psi (l(\theta^{-1}) (x, [\beta]))
\label{obsgroup}
\ee

On the other hand, the right action of $\pi_1(\M)$ in $\tilde \M$ defines the
{\bf \em right regular representation} $\R_r$ of $\pi_1(\M)$
in $l^2(\pi_1(\M))$ and a unitary representation
$ I \times \R_r(\eta)$ in $L^2(\tilde\M, d\mu)$,
\be
(I \times \R_r(\eta)) \, \psi (x, [\beta]) = \psi (r(\eta^{-1}) (x, [\beta])) \,  , 
\label{gaugegroup}
\ee
which, by eqs.\, (\ref{aMt1}), (\ref{UMt1}), commutes with $\tilde\pi_S(\A(M))$.

We recall that the left and right regular representations
$\R_l,\R_r$ of a discrete group $G$ in $l^2(G)$ (with basis $e_g, \, g \in G$) 
are unitarily equivalent and generate isomorphic
Von Neumann algebras ${\cal N}_l$, ${\cal N}_r$, which are the commutant one of the other,
${\cal N}'_l = {\cal N}_r$, ${\cal N}'_r = {\cal N}_l$.
Therefore, their centres coincide and give the central decomposition of both
$\R_l$ and $\R_r$ in $l^2(G)$.
Moreover, for any central projection $P$, the left and right representations in 
the corresponding space $P \, l^2(G)$ have $P \, e_{1} $, $1$ the identity in $G$,
as a cyclic vector and are defined by
complex conjugate matrix elements
$$ (P e_1, \, R_l(g) P e_1) = \overline {(P e_1, \, R_r(g) P e_1)} \, . $$

Keeping the same notation for $G = \pi_1(\M)$,
it follows from Theorem 2.9 and eqs.(\ref{UMt1}),(\ref{obsgroup}) that  
${\cal N}_l \cap {\cal N}'_l$ is the centre of the Von Neumann closure of the
observable algebra, $\tilde \pi_S (\A(\M))''$.
Since, by the irreducibility of the
Schroedinger representation, the algebra generated by $\tilde \pi_S (\alpha)$ and
$\tilde \pi_S (U(g_{\lambda v}))$, with $\theta(g,x) = 1$ \, a.e. in $x$,
is weakly dense in $L^2(\M)$),
$$\tilde\pi_S(\A(M))' = {\cal N}_r\, .$$
We have therefore:

\begin{Theorem}
In the representation $\tilde \pi_S(\A(\M))$ in
$$
L^2(\tilde \M, d\tilde\mu) \sim L^2(\M, d\mu) \times l^2(\pi_1(\M))
$$
the observable factors classifying the representations of $\A(\M)$ (Theorem 2.9)
are given (a.e.) by the left regular representation of $\pi_1(\M)$ in $l^2(\pi_1(\M))$,
eq.(\ref{obsgroup}).

\noindent
$\pi_1(\M)$ also
acts as a gauge group by its right regular representation,
eq.(\ref{gaugegroup}), which generates the commutant of $\tilde \pi_S(\A(\M))$.

\noindent
The observable and gauge representation of $\pi_1(\M)$ are unitarily equivalent.
The centres of the Von Neumann algebras generated by them
coincide and give the same reduction of $\tilde \pi_S(\A(\M))$.  
\end{Theorem}

\vspace{2mm}
\noindent 
e) {\em Completeness of the Dirac approach}

\vspace{1mm}

If (and only if)  $\pi_1(\M)$ is amenable, its regular representation contains \cite{Pier}
all its irreducible representations (in the weak, integral decomposition, sense if it is infinite).
In its reduction, as recalled above, central projectors gives rise to left (observable)
and right (gauge) representations $R$ and $\bar R$, defined by complex conjugate matrix elements
on a cyclic vector. This gives
\begin{Theorem}
If $\pi_1(\M)$ is amenable, the central decomposition of
$\tilde \pi_S (\A(\M))$ in $L^2 (\tilde \M, d\tilde \mu)$ is given by 
\begin{equation}
L^2(\tilde \M) = \sum_i L^2(\M) \times K^o_i  \times  K^g_i \, ,
\end{equation}
the sum ranging over all the irreducible representations $R_i$ of $\pi_1(\M)$.
$\pi_1(\M)$ acts as an observable group in $K^o_i$, with representation $R_i$,
and as a gauge group in $ K^g_i$, with a complex conjugate representation
$\bar R_i$.
The sum is replaced by an integral if $\pi_1(\M)$ is infinite,
with all the irreducible representations
appearing in the support of the corresponding measure.
\end{Theorem}

If $\pi_1(\M)$ is not amenable, 
Eqs.(\ref{aMt1}),(\ref{UMt1}) still allow for
the construction of all the irreducible LS representations of $\A(\M)$ on 
\emph {suitable, non $L^2$} Hilbert spaces of wavefunctions on $\tilde\M$.

In fact, given an irreducible unitary representation $\R$ of $\pi_1(\M)$
in a (automatically separable) Hilbert space $K$ and a non-zero vector $v \in K$ (cyclic
by irreducibility), a scalar product on the vector space $V$ of finite linear combinations
$\sum_g \lambda_g e_g$, $e_g \in \pi_1(\M)$  is defined by
\begin {equation}
  (e_g , e_h)_\R \equiv (\R(g) v, \R(h)v)_K
  \label{scalprod}
\end{equation}  
The left representation $\rho_l$ of $\pi_1(\M)$ in $V$, $ \rho_l(g) e_g \equiv e_{g\circ h} $,
preserves the above scalar product since
$$ (\rho_l(g) e_h, \rho_l(g) e_m)_\R =  ( \R(g \circ h) v, \R(g \circ m) v)_K =
( \R(h) v, \R(m) v)_K = (e_h, e_m)_\R \, . $$
It therefore extends to the Hilbert completion $\overline {V}_\R$ of $V$, where it
is unitarily equivalent to $\R$, the two representations
being given by the same expectations on the cyclic vectors $e_1$, $v$.

Then, on the completion $\H_\R$ of the space of complex functions $\psi(x,[\beta])$,
on $\tilde \M \sim \M \times \pi_1(\M)$, with finite support in the second argument and
scalar product
\begin{equation}
  (\psi(x,[\beta]), \chi(x,[\beta])) \equiv \int ( \overline {\psi(x,[\beta])},\chi(x,[\beta]))_\R \, d\mu(x)
  \label{prodR}
\end{equation}  
eqs.(\ref{aMt1}), (\ref{UMt1}), (\ref{thetadigamma}) define a representation of
$\A(\M)$ having $\R$ as associated representation of $\pi_1(\M)$.
By Theorem 2.9, we have therefore:
\begin{Proposition}
All the irreducible, locally normal, representations of $\A(\M)$ are unitarily equivalent
to representations, defined by eqs.\,(\ref{aMt1}), (\ref{UMt1}), (\ref{thetadigamma}), 
on Hilbert spaces of functions on $\tilde \M$,
with scalar product defined by eq.(\ref{prodR}).
\end{Proposition}  
If the corresponding representation of $\pi_1(\M)$ is finite dimensional, the scalar product
can be chosen so that the gauge group is represented in the resulting spaces.
In fact, in this case, 
eq.(\ref{scalprod}) can be substituted by 
\begin {equation}
  (e_g , e_h)_\R \equiv Tr_K (\R(e_g)^{-1} \R(e_h)) \, .
  \label{scalprodtr}
\end{equation}
The right action of $\pi_1(\M)$,
$$ \rho_r(e_g) e_h \equiv e_{h \circ g^{-1}} \, , $$
then defines a unitary representation of $\pi_1(\M)$
in $\bar V_\R$, and therefore in $\H_\R$,
commuting with the representation of $\A(\M)$ and complex conjugate to $\R$.

\end{document}